\documentclass{ieeeaccess}
\usepackage{cite}
\usepackage{amsmath,amssymb,amsfonts}
\usepackage{algorithmic}
\usepackage{graphicx}
\usepackage{textcomp}
\usepackage{float}
\usepackage{threeparttable}
\usepackage{refstyle}
\usepackage{amsmath}
\usepackage{footnote}
\usepackage{tablefootnote}
\usepackage{adjustbox}
\usepackage{subfig}
\usepackage{xcolor}
    \usepackage{caption,setspace}
\usepackage{textcomp}
\usepackage{multirow}
\def\BibTeX{{\rm B\kern-.05em{\sc i\kern-.025em b}\kern-.08em
    T\kern-.1667em\lower.7ex\hbox{E}\kern-.125emX}}
\begin{document}
\history{}
\doi{}

\title{Physical Multi-Layer Phantoms for Intra-Body 
Communications}
\author{\uppercase{Ahmed E. Khorshid}\authorrefmark{1,}\authorrefmark{2}, \IEEEmembership{Member, IEEE},
\uppercase{Ibrahim N. Alquaydheb\authorrefmark{1,}\authorrefmark{2},
}
\uppercase{Ahmed M. Eltawil\authorrefmark{1,}\authorrefmark{2}}, \IEEEmembership{Senior Member, IEEE}, \uppercase {and Fadi J. Kurdahi}.\authorrefmark{1,}\authorrefmark{2},
\IEEEmembership{Fellow, IEEE}}
\address[1]{Electrical Engineering and Computer Science Department, Henry Samueli School of Engineering, University of California, Irvine, CA}
\address[2]{Emails: \{khorshia, alquaydi, aeltawil, kurdahi @uci.edu\}}
\tfootnote{This work was supported in part by the U.S. National Institute of Justice under 2016-R2-CX-0014.}

\markboth
{Ahmed E. Khorshid \headeretal: Physical Multi-Layer Phantoms for Intra-Body 
Communications}
{Ahmed E. Khorshid \headeretal: Physical Multi-Layer Phantoms for Intra-Body 
Communications}

\begin{abstract}
This paper presents approaches to creating tissue mimicking materials that can be used as phantoms for evaluating the performance of Body Area Networks (BAN). The main goal of the paper is to describe a methodology to create a repeatable experimental BAN platform that can be customized depending on the BAN scenario under test. Comparisons between different material compositions and percentages are shown, along with the resulting electrical properties of each mixture over the frequency range of interest for intra-body communications; 100 KHz to 100 MHz. Test results on a composite multi-layer sample are presented confirming the efficacy of the proposed methodology. To date, this is the first paper that provides guidance on how to decide on concentration levels of ingredients, depending on the exact frequency range of operation, and the desired matched electrical characteristics (conductivity vs. permittivity), to create multi-layer phantoms for intra-body communication applications. 
\end{abstract}


\titlepgskip=-15pt

\maketitle

\section{Introduction}
\label{sec:introduction}

Wearable devices are rapidly being adopted as means of augmenting and improving health care services. In order to provide a cable-free biomedical monitoring system, new wireless technologies associated with sensor applications have been promoted as the next biomedical revolution, promising a significant improvement in the quality of health-care applications. Yet the size and power requirements of wireless sensors which are typically dominated by the RF section of the associated transceivers, has limited their adoption. To overcome such concerns, system architects proposed designing the system in a way that would allow more than one sensor to share the same wireless gateway, providing a distributed solution, with less power consumption. That approach paved the way for adopting Intra-Body Communication (IBC) systems where data transmission is carried out through the body (mostly skin layers), rather than through air \cite{zimmerman1995personal,handa1997very}. Sensors and actuators inter-communicate through skin, relaying messages to a centralized wireless hub that could be a smart watch for instance. This emerging technology would ultimately lead to Body Area Networks (BANs) that operate at extremely low power, with minimal foot print by replacing expensive, power consuming Radio Frequency (RF) front ends, for each individual node with simpler interfaces. Furthermore, while the skin operates as an interference channel for the communicating nodes, it is relatively protected from the higher levels of interference expected when broadcasting via an air interface.  
\par
Initial work focused on evaluating the suitability of the human body as a communication medium. Research efforts were first directed towards modeling the gain/attenuation profile of the body channel versus frequency and simulating its behavior using different software tools. The goal of this characterization will was to identify the  optimum frequency range for IBC as well as the frequency range at which the body's attenuation to the signal propagation would be minimal, thus minimizing the power needed for transmission. Factors affecting such profile were also considered, such as, type, shape and size of the electrodes used, distance between the transmitter and the receiver, biological parameters of the human body and the environmental conditions \cite{handa1997very,ruiz2006propagation,hachisuka2006simplified,wegmueller2007attempt,wegmueller2010signal,cho2007human,callejon2012distributed,kibret2014investigation,phantommodel}. However, simulation results alone are not sufficient and must be verified through comparison with experimental data from measurements that are carried out on real subjects. Carrying out experiments on real subjects is a tedious process that requires long and complicated procedures, mainly to ensure the safety of the subjects under test, especially for emerging technologies. The above facts, together with the urgent need for experimental data for verifying the proposed channel models, encouraged researchers to adopt the idea of using phantoms that mimic the characteristics of human tissue including thermal, physical and electrical characteristics, in their experimental routines. Phantoms are widely used in the medical and biological studies as substitutes for animals in experiments, lending themselves readily for BAN research. 
\par
The main contribution of this paper can be summarized as follows:
\begin{enumerate}
\item Describing a methodology to create \emph{multilayer} phantoms that mimic the electrical properties of the body tissues for low frequency applications such as the IBC, while capturing the effect of the multiple layers in the human body, namely, skin, fat, muscle, cortical-bone and bone-marrow.
\item Providing experimental measurement results for different prepared samples over the frequency range from 100 KHz till 100 MHz.
\end{enumerate}

In this paper, the following subsection presents a comparison between the different data carriers that are considered for intra-body communications, showing why Electro-Magnetic (EM) waves are the best candidate. Key features that should be captured by phantoms are presented and discussed. Next, phantoms are introduced, and two different methods for preparing phantoms suitable for IBC applications are presented. Experimental procedures and setup for preparing different materials constituting the phantom are detailed. Measurement results are discussed, and properties of the prepared samples are compared with those of the tissues of concern. Finally, a composite model is proposed, wherein results for the experimental measurements are shown proving the efficacy of the methodology in producing multilayer phantoms.

\section{Optimum data carrier for intra-body communication}
The first concern for IBC systems is determining the optimum medium that would serve as the best data carrier for intra-body communication applications. In \cite{khorshid2016optimum}, the authors presented a comparative study between the main potential means of data carriers, namely using electro-magnetic waves, ultrasonic waves and magnetic coupling. Properties of each carrier were considered with respect to their propagation through the different body tissues. It was shown in \cite{khorshid2016optimum} that EM waves possess better properties that can support BAN requirements compared to ultrasonic waves as EM waves experience much less attenuation and delay when traveling through the body. Thus, EM was selected as the data carrier of choice for IBC applications.
% Then a more in-depth comparison was shown between EM waves and ultrasonic waves concerning the attenuation and delay that each wave would experience when propagating through the human body, within the frequency range of interest for intra-body communication applications. Such characteristics are strongly dependant on the electrical properties of the tissues, where for the EM, attenuation was calculated according to the following equation:
% \begin{equation}
% \alpha=\frac{\omega}{2}\sqrt[]{\mu\epsilon(\omega)}\tan(\delta(\omega))
% \end{equation}
% where $\alpha$ is the attenuation (dB/cm), $\omega$ frequency (rad/s), $\mu$ permeability, $\epsilon(\omega)$ frequency dependant permittivity and $\delta(\omega)$ is the loss tangent, and these electrical properties were all calculated for each tissue \cite{ruiz2006propagation}.
\par
Intra-body communications using EM waves as data carriers can be categorized into two main types; capacitive coupling (near field coupling method) and galvanic coupling. In capacitive coupling, only the signal electrodes of the transmitter and the receiver are attached to the body while the ground (GND) electrodes are left floating in the air. The conductive body forms the forward path while the signal loop is closed through the capacitive return path between the transmitter and the receiver GND electrodes. The second approach, which depends on the galvanic coupling principle, uses a pair of electrodes for both the transmitter and the receiver to propagate the electromagnetic wave. The signal is applied over two coupler electrodes and received by two detector electrodes. The coupler establishes a modulated electrical field, which is sensed by the detector. Therefore, a signal transfer is established between the coupler and detector units by coupling minute signal currents into the human body. In both approaches, it has been shown that the attenuation of the body channel can be much lower than that of the air channel at frequencies up to 100 MHz \cite{ruiz2006propagation}.
\section{Phantoms}
Phantoms are physical models that simulate certain characteristics of the biological tissues they represent. Phantoms have been used extensively in the medical field \cite{artificialphantom}. Historically known imaging phantoms were first introduced as objects for evaluating the performance of imaging devices. Phantoms then underwent various improvements, mimicking biological characteristics more accurately, where they proved to be useful solutions for experimentation at the early investigative stages prior to working with living subjects or cadavers. There are various classifications for phantoms, the most commonly used is according to the final state of the phantom; solid (dry or wet), gel or liquid \cite{stableflex,scattering,augustine2009electromagnetic,guy1971analyses,ito2001development,kobayashi1993dry,tamura1997dry,nikawa1996soft,chang2000conductive}. Finally, phantoms provide a stable and more controllable experimental setup/platform that is hard to realize using living subjects. 
\subsection{Phantoms for IBC}
Several trials were reported in literature for the use of phantoms for IBC applications as a stable and easy to control, yet accurate experimental setup \cite{presenttrends,wirelesssurvey}. Liquid phantoms were usually adopted in these trials, being the easiest to prepare. In \cite{hachisuka2005intra}, the authors used an insulator (polyvinyl chloride bag) containing conductive liquid (salt water) to model the human arm as a cylinder. In \cite{tang2011channel}, the authors used a liquid phantom as well that consists of 0.45$\%$ NaCl and 2 gallons of water filled in a plastic container, yielding a solution of conductivity $\sim$0.52 S/m at 13MHz and $\sim$1S/m at approximately 900MHz. In \cite{ishida2013signal}, the authors used a phantom that is a gel material with a conductivity of 0.59 S/m at 6.75 MHz and packed by a plastic sheet. {In \cite{136}, the authors constructed a circular phantom consisting of two homo-centric sections each filled with a different substance; water with a given sodium concentration and agar. Agar was used to emulate the skin and a saline solution accounted for the interstitial fluid and muscle. In \cite{139}, a  semi-cylindrical container was proposed formed by two homo-centric layers composed of different chemical compounds emulating the skin and muscle. In \cite{42}, the authors proposed a solid phantom in the form of a rectangular parallelepiped with a relative permittivity of 81 and a conductivity of 0.062 S/m. Wegmueller et al. showed in \cite{51}, an ellipsoidal phantom for the simulation of a cross-section of the torso, filled with a muscle simulating fluid that emulates the conductivity of muscle at 27MHz. 

\par
A main drawback for the above mentioned trials is that the proposed phantoms all considered the arm as one homogeneous layer; using a single material to represent the whole arm. Such approach results in:
\begin{enumerate}
\item Neglecting the biological and physical nature of the arm, mainly the dielectric properties of each of the main five tissues, and over the whole range of frequencies of interest (instead of just reporting such values at one or two discrete frequencies).
\item Considering the arm as one single layer also neglects the interaction between one layer (tissue) and the other, which accordingly neglects and eliminates other important facts of how the signal would diffuse from one layer to the other, propagation of the signal in different layers, etc.
\item Inaccurate representation of the arm's geometry; dimensions of the arm and thickness of each layer (which has a considerable impact on the overall results and performance for IBC) are almost totally neglected.
\end{enumerate}

For the above reasons, it became clear that if phantoms are to serve as stable, controllable, accurate and reliable testing setup for IBC, then more elaborate and detailed phantoms need to be used. In prior work, the authors \cite{khorshid2015intra} proposed an accurate circuit model of the human arm as an IBC channel. In that model, the arm was simplified to the five main layers previously mentioned; muscle, fat, skin, cortical bone and bone marrow. Since the gain profile obtained using the model showed very good match with experimental results previously reported in the literature, the authors opted to follow the same approach in constructing the proposed phantom. Two different methods were studied to construct the samples that would satisfy our goal.
\subsection{Oil Phantoms}
   \begin{figure}[]
       \centering
 \includegraphics[width=2.2in,scale=1]{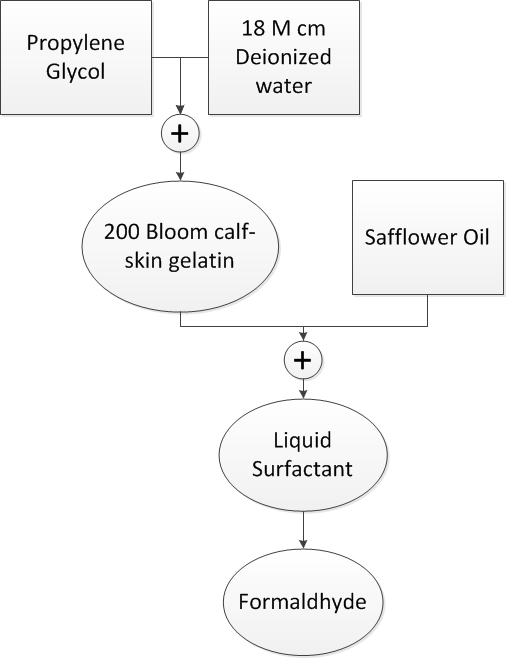}
 \caption{Elements used in preparing the oil phantom\cite{madsen2006anthropomorphic}.}
 \label{fig1_oil_only_fig1}
    \end{figure}
In \cite{madsen2006anthropomorphic}, equivalent anthropomorphic breast phantoms were constructed for use in ultrasound elastography and magnetic resonance (MR) elastography. The proposed method seemed to possess a potential for yielding materials with dielectric properties that can be varied through varying the ratios of the elements used in the preparation of each sample. The main elements/components used for preparing the materials are shown, with the order at which each element is added throughout the preparation process for each material as shown in Figure \ref{fig1_oil_only_fig1}. 
The liquid aqueous gelatin and the safflower oil are the main two components of the mixture. By varying the ratios of these two components each time, a new material with different dielectric properties is generated. Each resultant sample is identified by the percentage of oil to the total final sample weight; for instance, a 50$\%$ sample has safflower oil at 50$\%$ of the weight of the final product. The main steps for preparing each sample are as follows \cite{madsen2006anthropomorphic}:
\\
1. In a beaker, prepare a room temperature solution of propylene glycol and 18 megohm-cm doubly de-ionized water.
\\
2. Slowly add, while stirring, 200 bloom calf-skin gelatin so that no clumping occurs and a uniform "slurry" results.
\\
3. Cover the beaker with a plastic food wrap held in place with a rubber band. Punch a small hole or slit in the plastic wrap so that the gas pressure above the slurry during heating remains at atmospheric pressure.
\\
4. Place the beaker in a larger container of hot water so that the level of the hot water is at or above the top of the gelatin slurry in the beaker.
\\
5. Heat the water until the gelatin temperature rises to about 90$^{\circ}$C and becomes transparent. Remove any bubbles at the meniscus. The transparent hot gelatin is referred to below as molten gelatin.
\\
6. Remove the beaker of molten gelatin from the hot water bath and immerse it partially in a cold water bath. Cool the molten gelatin, while stirring, to 50$^{\circ}$C and remove it from the cold water bath.
\\
7. While cooling the molten gelatin in step 6, heat the safflower oil to 50$^{\circ}$C.
\\
8. Add the molten gelatin to the 50$^{\circ}$C safflower oil and mix vigorously with a tablespoon that is bent at right angles near the bowl of the spoon. During mixing, keep the bowl of the spoon beneath the surface and moving about a horizontal axis, thus minimizing disturbance to the surface of the mixture. 
\\
9. Add the liquid surfactant and continue the stirring motion until the emulsion is nearly white and a separation of oil does not occur when stirring is stopped.
\\
10. Cool in the cold water bath to 40$^{\circ}$C and slowly add with stirring formalin.
\\
11. Continue cooling the emulsion to about 34$^{\circ}$C and pour into molds.
\\
12. Allow at least 8 hours for cross-linking of the gelatin by formaldehyde to occur before removing the phantom component from its mold.
\subsection{Oil-Kerosene Phantoms}
 \begin{figure}[]
       \centering
 \includegraphics[width=2.8in,scale=1]{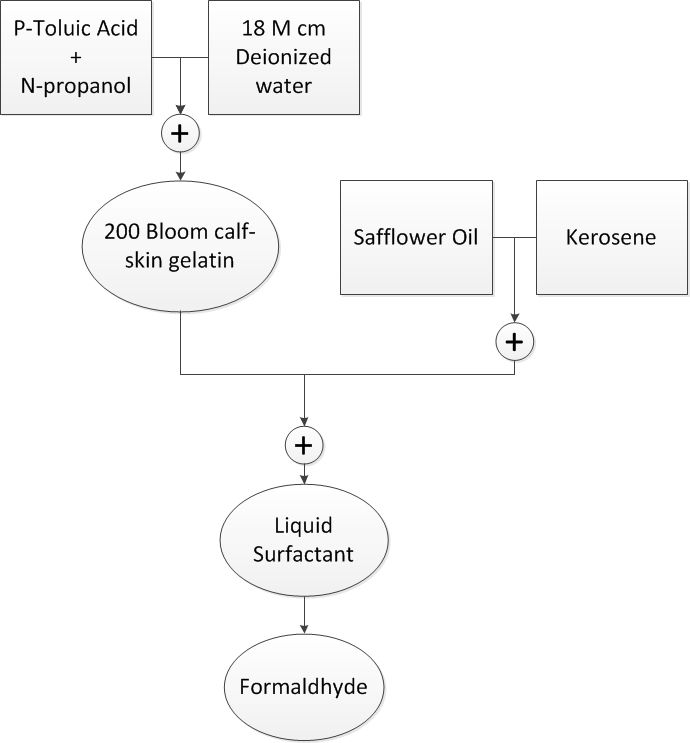}
 \caption{Elements used in preparing the oil-kerosene samples \cite{kanda2004formulation}.}
 \label{fig2_oil_and_kerosene_only_fig2}
    \end{figure} 
In \cite{kanda2004formulation}, tissues mimicking phantom materials were proposed for microwave applications, where satisfactory results were reported for different samples, yet for a higher frequency range than the one we are interested in. To be able to use these materials within the frequency range of interest for IBC applications (100 KHz till 100 MHz), some modifications were applied. The main elements used in preparing the samples are shown in Figure \ref{fig2_oil_and_kerosene_only_fig2}.
  \begin{figure}[]%
      \centering
      \subfloat[]{{\includegraphics[height=2.5in, width=3.28in]{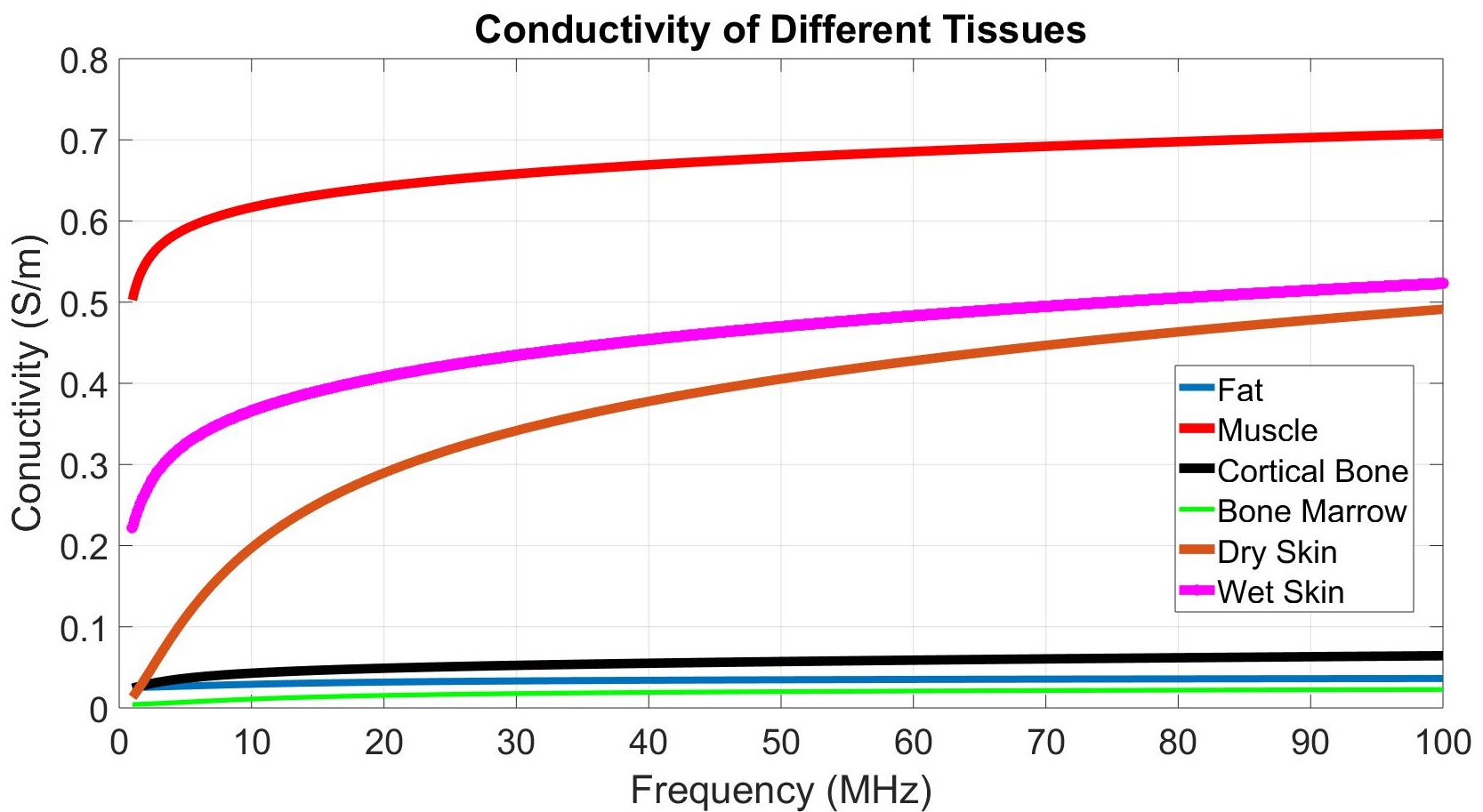} }}%
    \qquad  
      \subfloat[]{{\includegraphics[height=2.5in, width=3.28in]{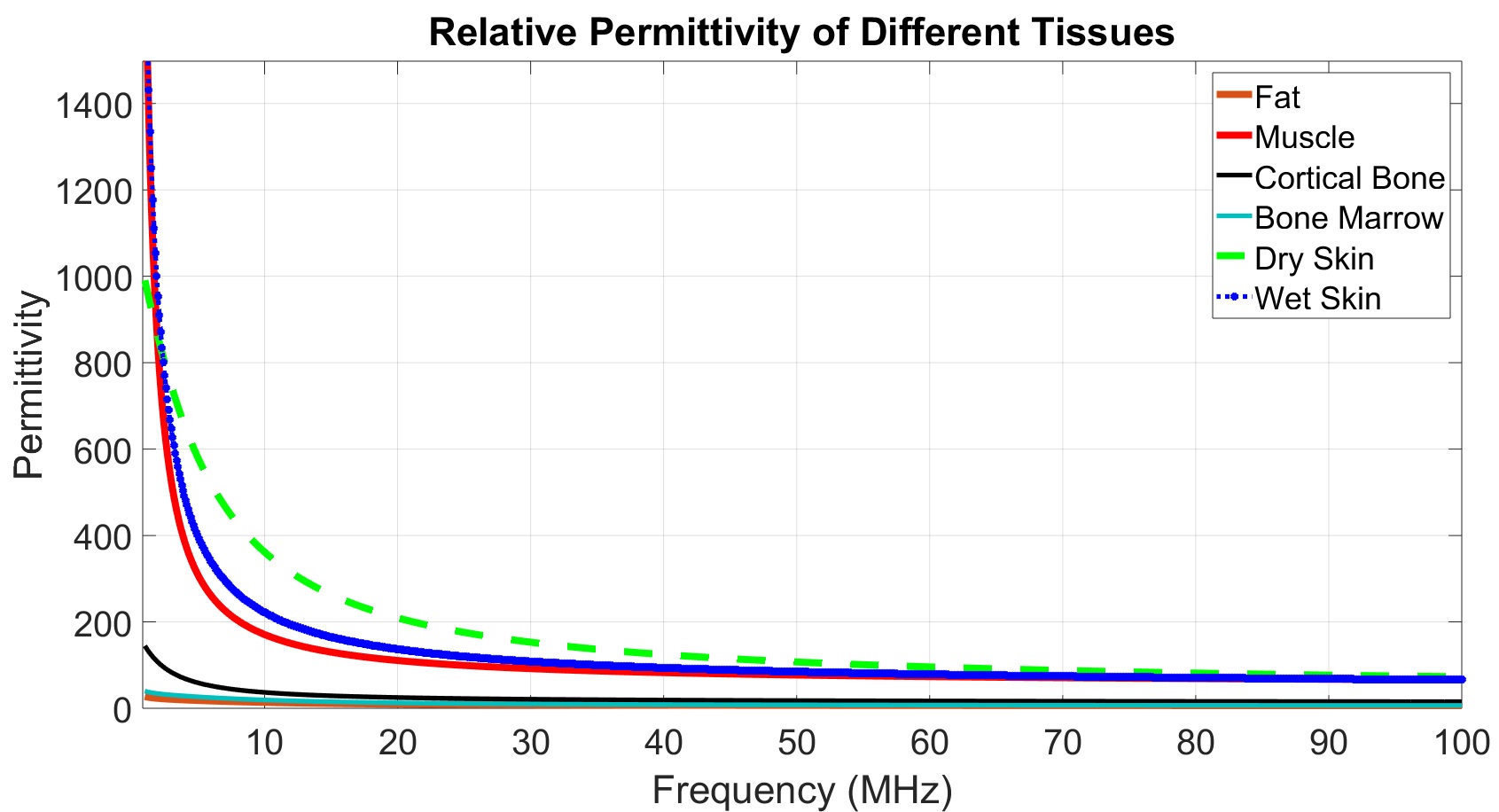} }}%
      \caption{Electrical properties of the main five tissues (skin, muscle, fat,
  cortical bone and bone marrow) using the experimental measurement values reported in \cite{gabriel1996dielectric,gabriel2}.}%
      \label{fig3_dielectric_properties}%
  \end{figure}
  \begin{figure}[]%
      \centering
      \subfloat[]{{\includegraphics[height=2.5in,width=3.28in]{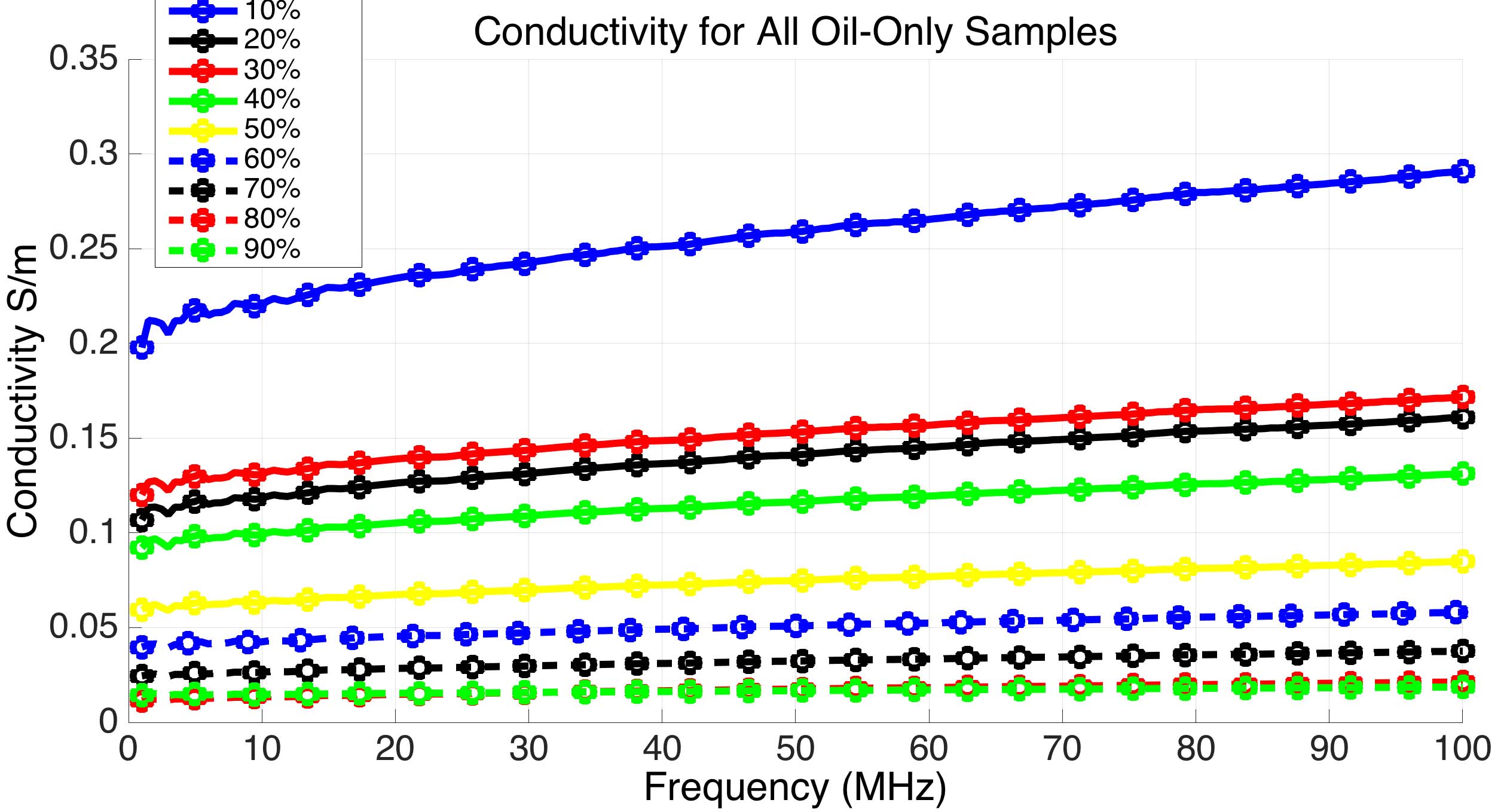} }}%
    \qquad  
      \subfloat[]{{\includegraphics[height=2.5in,width=3.28in]{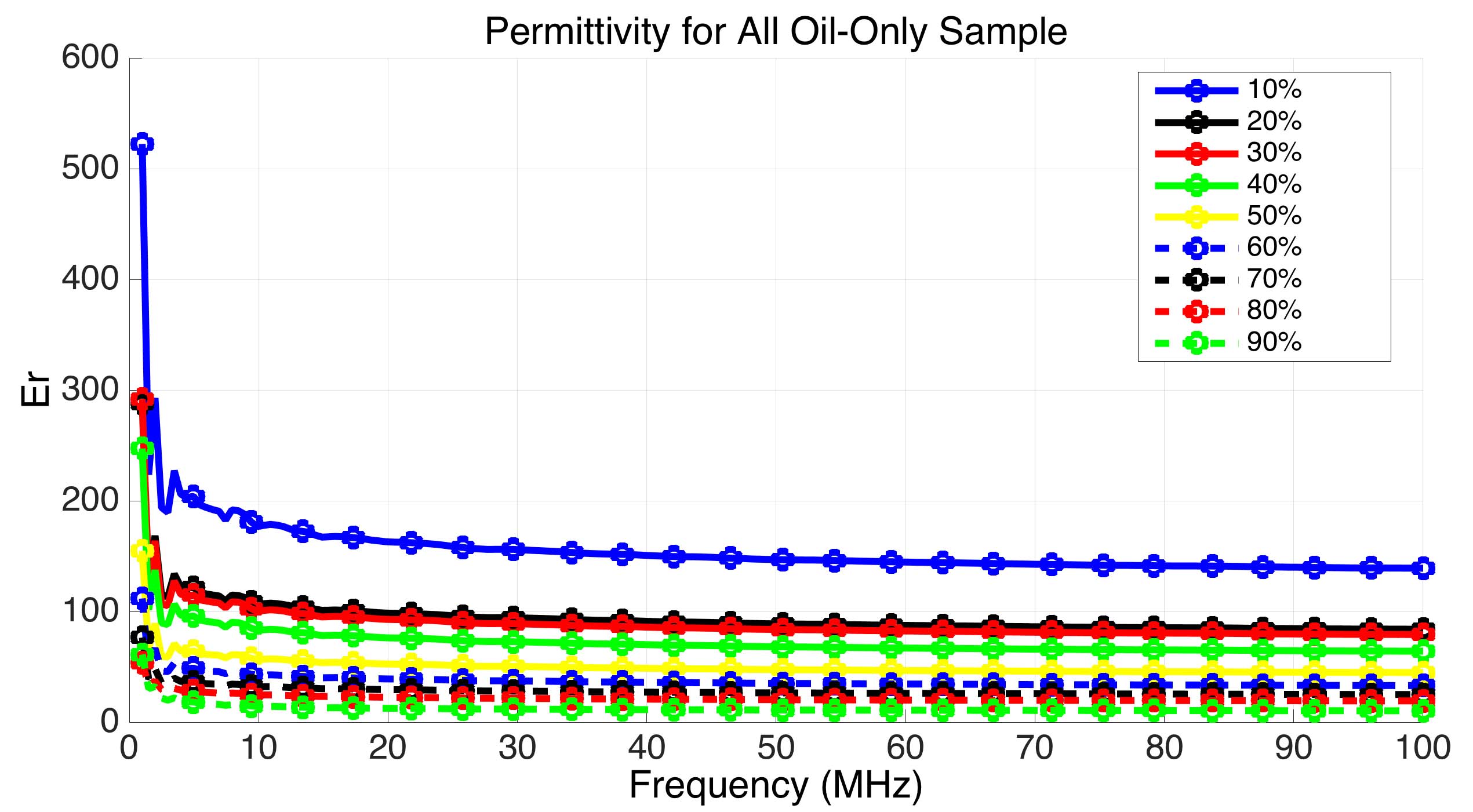} }}%
      \caption{(a) Measured conductivity for the Oil-only samples for the different Oil concentrations. (b) Measured permittivity.}%
      \label{fig4_dielectric_properties_oil_only}%
  \end{figure}
\par
The main difference between this method and the previous one is that the oil solution in this case would be an equal mix of kerosene and safflower oil. The preparation steps for these materials are very similar to those of the Oil Phantom mentioned earlier, and can be found in more details in \cite{kanda2004formulation}. Another important advantage to the two methods being adopted for preparing the phantom materials is that they can be employed in a heterogeneous configuration (samples with different concentrations being constructed side by side) without change in geometry or dielectric properties of any of them, a feature that is needed to construct a multi-layer phantom structure.
\begin{table*}[]
\centering
\caption{WEIGHTS OF ELEMENTS USED IN PREPARING THE OIL PHANTOM SAMPLES}
\label{table1_oil_only}
\begin{adjustbox}{max width= 34in}
\begin{tabular}{|l|l|l|l|l|l|l|l|l|l|}
\hline
\textbf{Material\textbackslash Sample Percentage} & \textbf{10\%} & \textbf{20\%} & \textbf{30\%} & \textbf{40\%} & \textbf{50\%} & \textbf{60\%} & \textbf{70\%} & \textbf{80\%} & \textbf{90\%} \\ \hline
Propylene Glycol(units)                           & 10.5          & 10.5          & 10.5          & 10.5          & 10.5          & 10.5          & 10.5          & 10.5          & 10.5          \\ \hline
De-ionized water(units)                           & 169           & 169           & 169           & 169           & 169           & 169           & 169           & 169           & 169           \\ \hline
Gelatin(units)                                    & 26.95         & 26.95         & 26.95         & 26.95         & 26.95         & 26.95         & 26.95         & 26.95         & 26.95         \\ \hline
Safflower Oil(units)                              & 19.4          & 43.75         & 75            & 116.7         & 175           & 262.5         & 408.3         & 700           & 1575          \\ \hline
Ultra Ivory(units)                                & 0.2314        & 0.48125       & 0.825         & 1.2837        & 1.925         & 2.8875        & 4.4913        & 7.7           & 17.325        \\ \hline
Formalin(units)                                   & 1.323         & 1.323         & 1.323         & 1.323         & 1.323         & 1.323         & 1.323         & 1.323         & 1.323         \\ \hline
\end{tabular}
\end{adjustbox}
\end{table*}
\begin{table*}[]
\centering
\caption{WEIGHTS OF ELEMENTS USED IN PREPARING THE OIL-KEROSENE PHANTOM SAMPLES}
\label{table2_oil_kerosene}
\begin{tabular}{|l|l|l|l|l|l|l|l|l|l|}
\hline
\textbf{Material\textbackslash Sample Percentage } & \textbf{10\% \,} & \textbf{20\% \enspace } & \textbf{30\% } & \textbf{40\% \enspace } & \textbf{50\% } & \textbf{60\% \enspace  } & \textbf{70\% \enspace } & \textbf{80\% } & \textbf{90\%  \enspace \enspace   } \\ \hline
P-Toluic acid (g)                                 & 0.2           & 0.2           & 0.2           & 0.2           & 0.2           & 0.2           & 0.2           & 0.2           & 0.2           \\ \hline
De-ionized water (ml)                             & 190           & 190           & 190           & 190           & 190           & 190           & 190           & 190           & 190           \\ \hline
N-Propanol (ml)                                   & 10            & 10            & 10            & 10            & 10            & 10            & 10            & 10            & 10            \\ \hline
Gelatin (g)                                       & 34            & 34            & 34            & 34            & 34            & 34            & 34            & 34            & 34            \\ \hline
Oil-kerosene (ml)                                 & 22.2          & 50            & 85            & 133.3         & 200           & 300           & 466           & 800           & 1800          \\ \hline
Ultra Ivory (g)                                   & 1.26          & 2.8           & 4.76          & 7.46          & 11.2          & 13            & 15            & 17            & 20            \\ \hline
Formalin (g)                                      & 2.16          & 2.16          & 2.16          & 2.16          & 2.16          & 2.16          & 2.16          & 2.16          & 2.16          \\ \hline
\end{tabular}
\end{table*}
\section{EXPERIMENTAL RESULTS}
To investigate the potential of both methods - i.e. prepare samples that would have the closest dielectric properties to those of the five body tissues we are trying to mimic- nine different samples of each method were prepared, where the oil (or oil-kerosene solution) percentage was varied from 10\% of the sample weight to 90\%, with a 10\% step each time. The exact weights and percentages of the elements used in preparing the samples are given in Table \ref{table1_oil_only} for the oil phantoms and Table \ref{table2_oil_kerosene} for the oil-kerosene ones.
\par
After pouring the prepared samples in their molds and allowing time to mature (at least for 5 days for the formaldehyde cross-linking of gelatin to be completed \cite{kanda2004formulation}), the dielectric properties were measured. An HP Agilent 4291B \cite{keysight_tech} impedance analyzer was used to perform the measurements. Both the real part and the imaginary part of the complex permittivity are measured using the device. Results are then processed using MATLAB software to compute both the permittivity and conductivity of each sample and compare them with those of the five body tissues. The 16453A dielectric material test fixture was attached to the 4291B impedance analyzer, where the function of this fixture is to obtain accurate dielectric constant and loss tangent measurements through employing the parallel plate method, which sandwiches the material between two electrodes to form a capacitor. To be able to use this fixture, the thickness of the sample to be tested should not exceed 3 mm. For each prepared material, at least three different test samples from different locations across each mold were used to make sure that the results are not affected by the position of the sample in the mold.
Results are then averaged over the collected readings to compute the dielectric properties for each prepared material. To investigate the properties of the prepared materials over time, measurements were repeated after 8 weeks to study the change in the properties of the material over time. 
% One issue was that for the 90\% Oil-Kerosene case the final sample was still in a semi-liquid state due to the high concentration of the oil solution, making it very slippery and fragile to be tested using the available setup.
\section{RESULTS}
Using the experimental measurements provided in \cite{gabriel1996dielectric,gabriel2}, the dielectric properties (conductivity and permittivity) for the main five tissues that are the scope of this study, skin, muscle, fat, cortical bone and bone marrow, are plotted in Figure \ref{fig3_dielectric_properties}.
\par
The dielectric properties of the samples are plotted in Figure \ref{fig4_dielectric_properties_oil_only} (for oil only samples) and Figure \ref{fig5_dielectric_properties_oil_kerosene} (for oil and kerosene samples). The figures show that as the oil or oil-kerosene concentration increase, both the conductivity and permittivity of the prepared sample decrease, which is expected as oil in general has more of an insulator properties (weak electrical conductivity, yet better heat conductivity). Comparing such results, with the tissue properties shown in Figure \ref{fig3_dielectric_properties}, it can be observed that both follow the same behavior with respect to frequency; conductivity increases with frequency, while permittivity experiences a sharp fall at lower frequencies and then almost saturates at higher frequency values. Comparing the electrical properties plotted in Figure \ref{fig3_dielectric_properties}  with those plotted for the different samples in Figure \ref{fig4_dielectric_properties_oil_only} \& \ref{fig5_dielectric_properties_oil_kerosene}, the accurate oil percentage can thus be selected depending upon which specific tissue needs to be mimicked, as well as which electrical property is of more concern (conductivity or permittivity), and the frequency range of interest. For instance, samples with low oil solution concentration can be utilized for mimicking tissue with high dielectric permittivity, while samples with high oil solution concentration can be used to mimic tissues with low conductivity like cortical bone, as shown in Figures
\begin{figure}[H]%
      \centering
      \vspace{-0.1in}
      \subfloat[]{{\includegraphics[height=2.5in,width=3.28in]{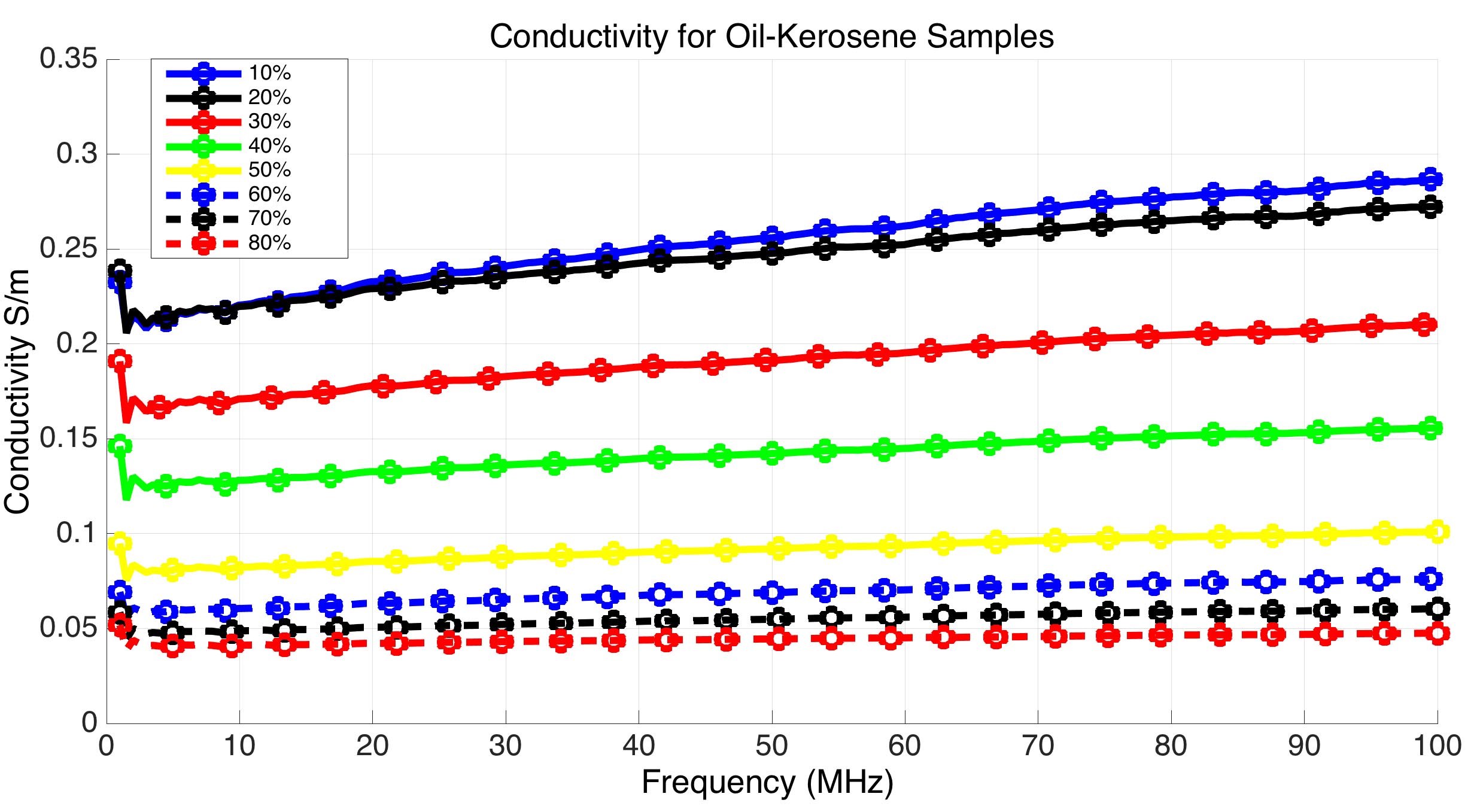} }}%
    \qquad  
      \subfloat[]{{\includegraphics[height=2.5in,width=3.28in]{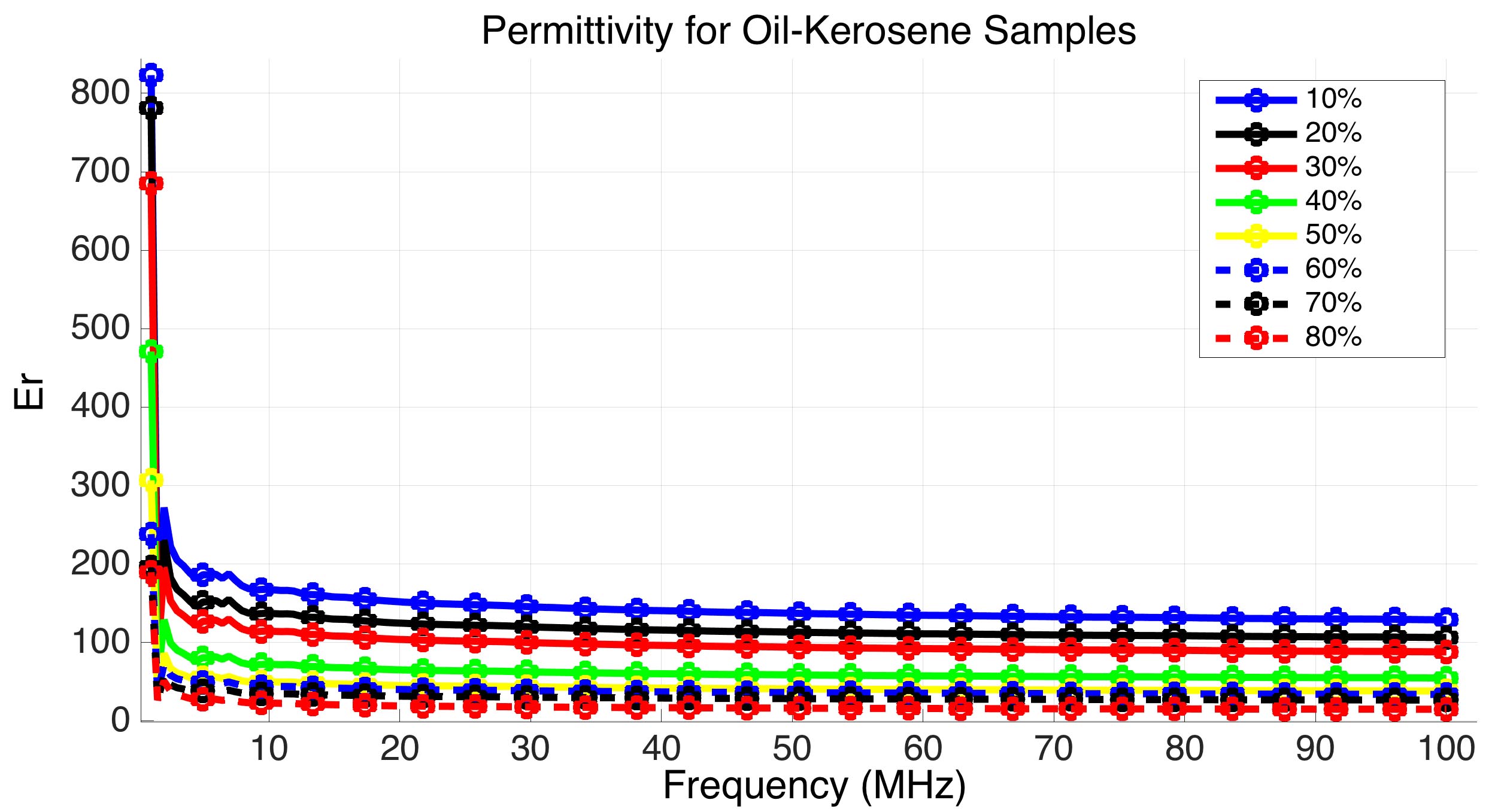} }}%
      \caption{(a) Measured conductivity for the Oil-Kerosene samples for the different Oil-kerosene to the gelatin solution concentrations. (b) Measured permittivity.}%
      \label{fig5_dielectric_properties_oil_kerosene}%
      \vspace{-0.1in}
  \end{figure} \ref{fig6_a_different tissues} and \ref{fig6_eighty_percent}, where the 80\% oil-kerosene solution sample perfectly mimics the permittivity characteristics of the cortical bone, specially for the frequency range between 30 MHz to 100 MHz. 

As shown in Figures \ref{fig6_a_different tissues} and \ref{fig6_eighty_percent}, the electrical properties of different tissues can be matched with samples of certain formulations and concentrations, depending on the electrical property of interest (whether conductivity or permittivity is more of concern), and the range of frequency in which the IBC application will operate within. It is important to note that most IBC applications utilize less than 1 MHz of bandwidth \cite{presenttrends}, due to the nature of medical applications, that typically require low bit rates. A summary of such results; best matching samples (samples that shows less than 10 \% matching error) with respect to different tissues, regarding conductivity and permittivity, for different frequency ranges (Fmin is the minimum frequency and Fmax is the maximum frequency in MHz defining the band over which the matching error is below 10\% ) within the IBC application band, is provided in Table 3.
 \begin{figure}[H]%
      \centering
      \vspace{-0.1in}
      \subfloat[]{{\includegraphics[height=2.6in,width=3.28in]{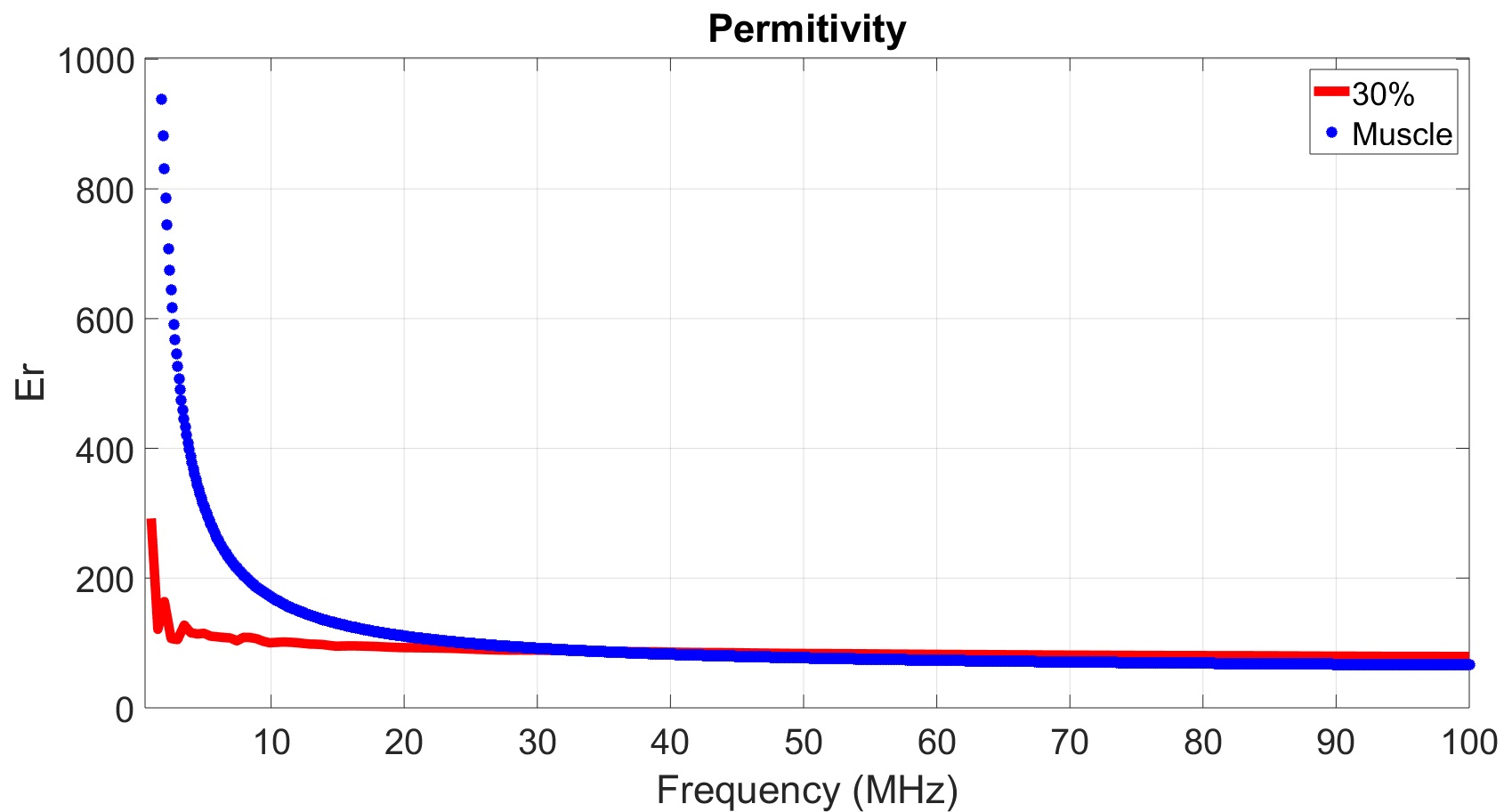} }}%
      \qquad  
      \subfloat[]{{\includegraphics[height=2.6in,width=3.5in]{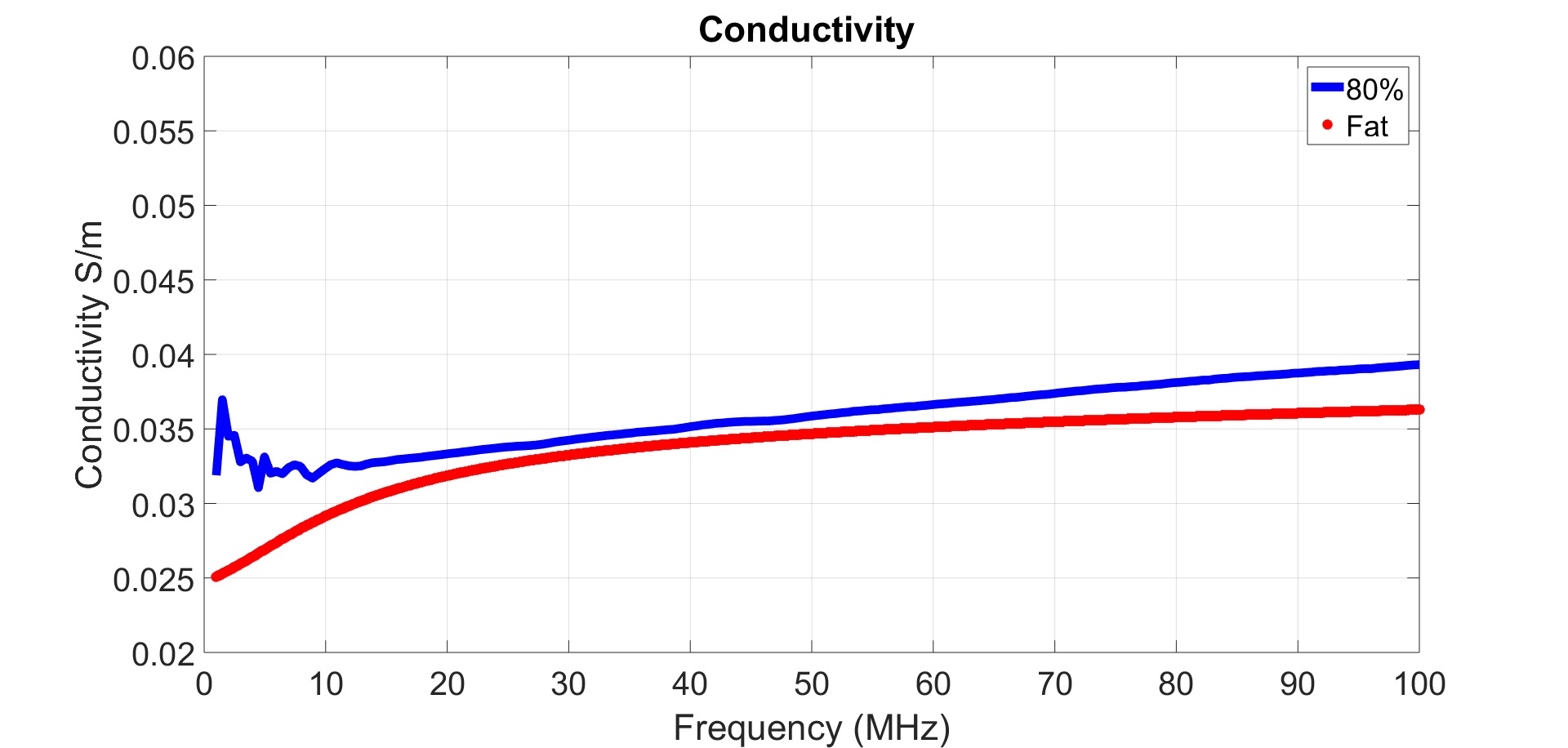} }}%
    \qquad  
      \subfloat[]{{\includegraphics[height=2.6in,width=3.28in]{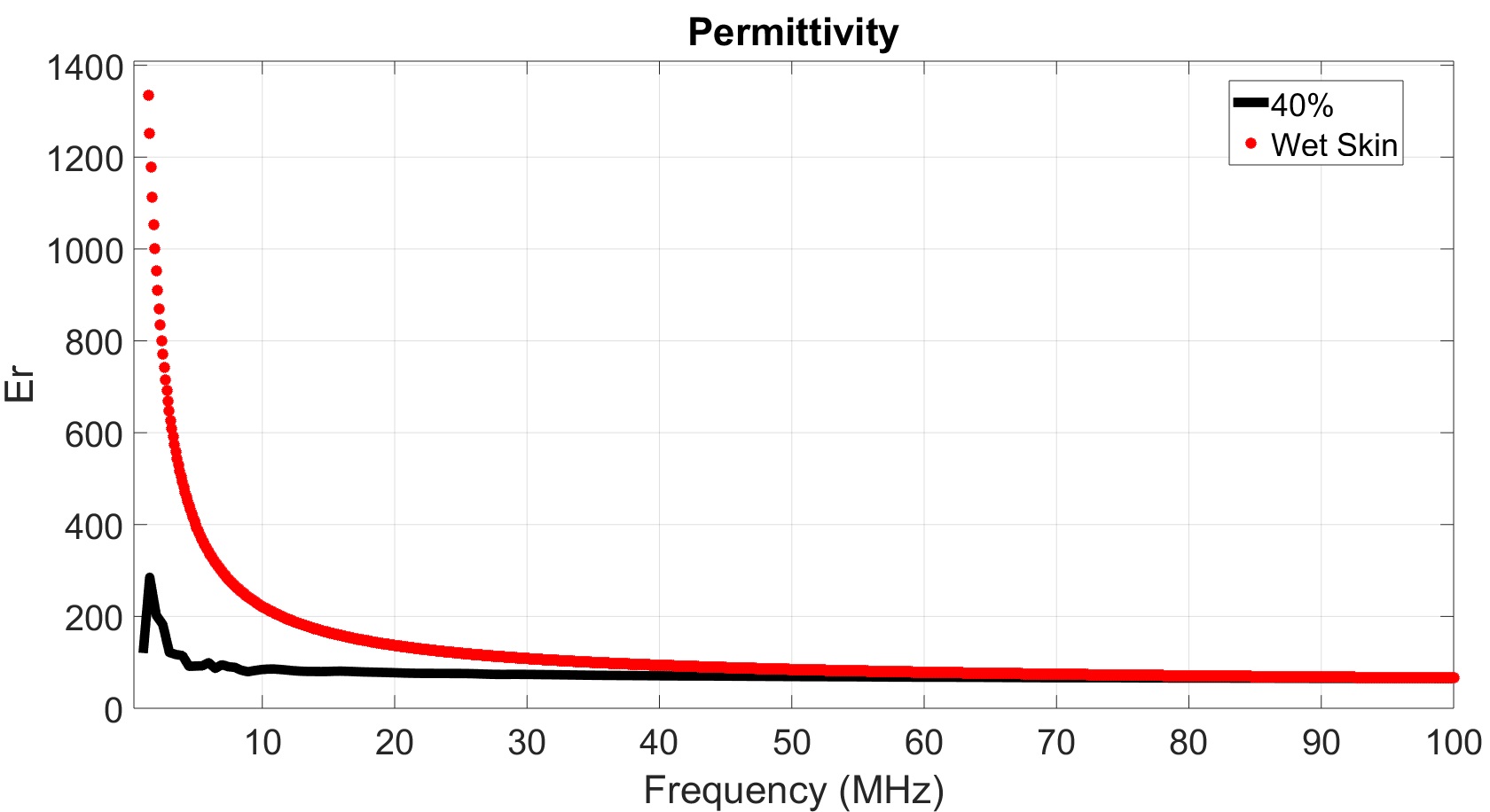} }}% 
      \caption{ An example for matching the samples with certain solution concentration with the tissue of concern within certain frequency bands: (a) 30\% oil-kerosene solution shows accurate match with the muscle tissue,from the permittivity point of view, for frequencies above 30 MHz, (b) 80\% oil-kerosene solution shows accurate match with fat tissue,from the conductivity point of view, for frequencies above 10 MHz,(c) 40\% oil-only solution for the skin (wet) tissue, regarding the permittivity, for frequencies above 40 MHz.  }%
      \label{fig6_a_different tissues}%
      \vspace{-0.19in}
  \end{figure}
 \begin{figure}[H]%
      \centering
      \subfloat[]{{\includegraphics[height=2.6in,width=3.28in]{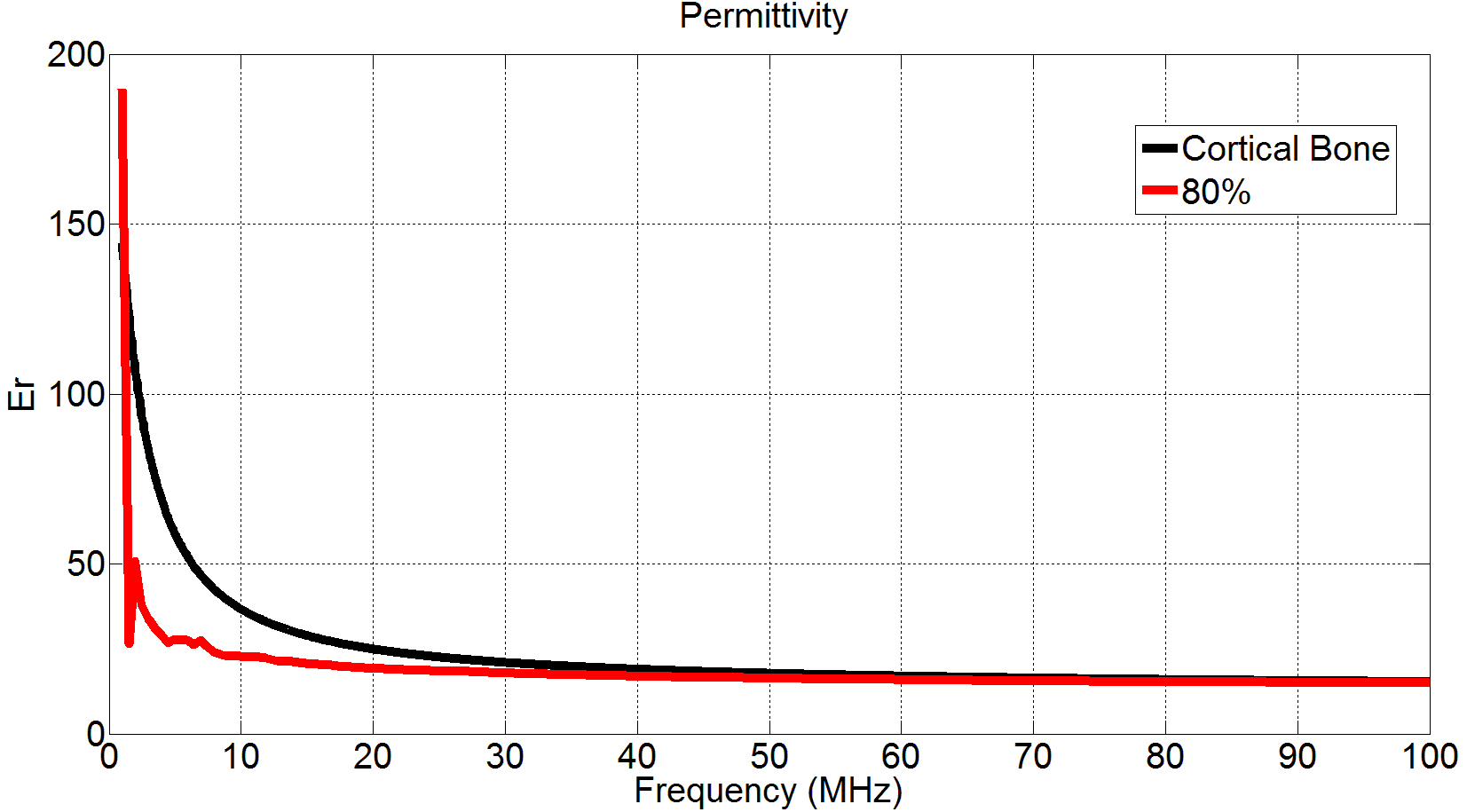} }}%
       \qquad  
     \subfloat[]{{\includegraphics[height=2.6in,width=3.28in]{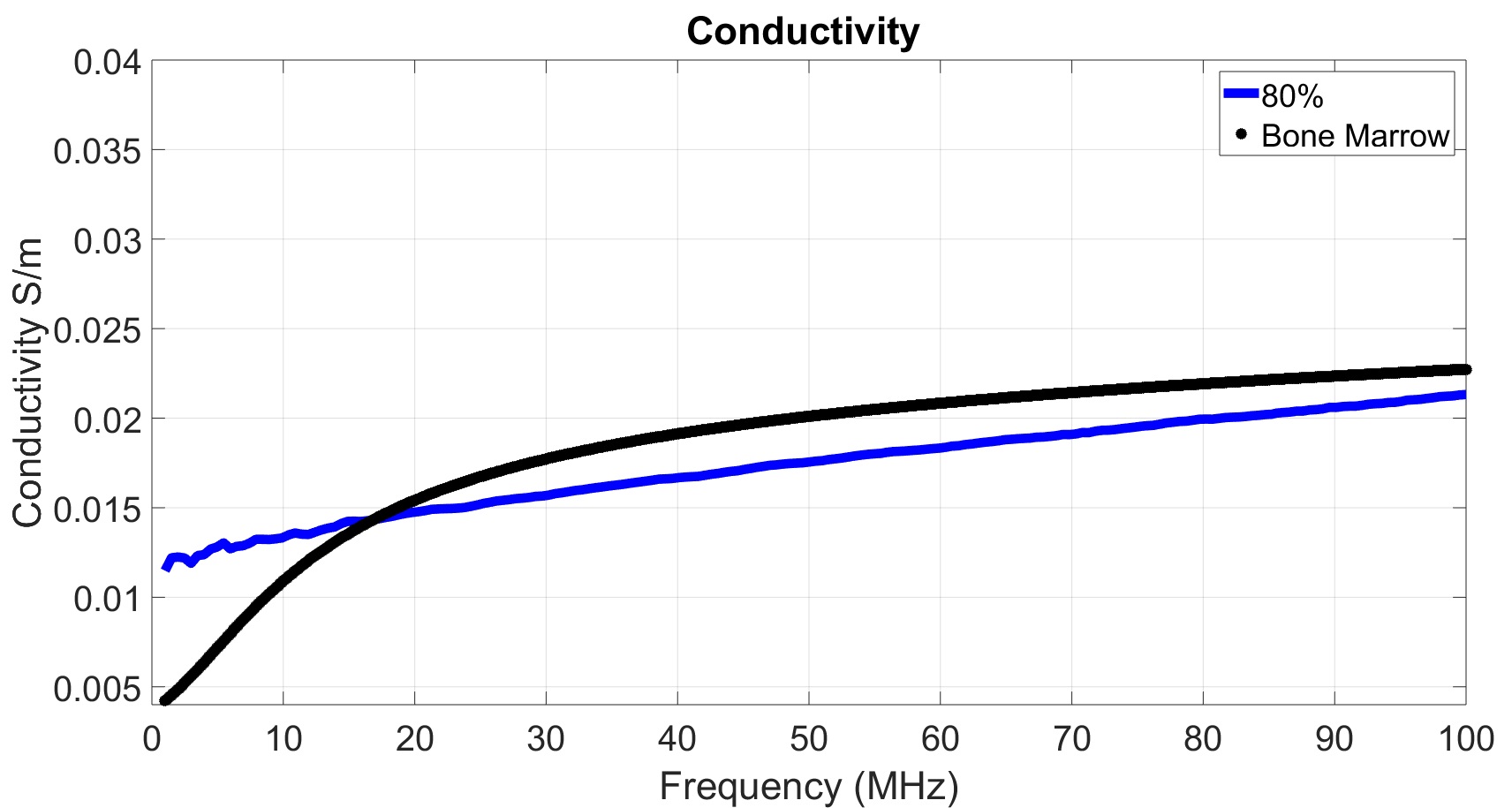} }}
      \qquad  
      \subfloat[]{{\includegraphics[height=2.6in,width=3.28in]{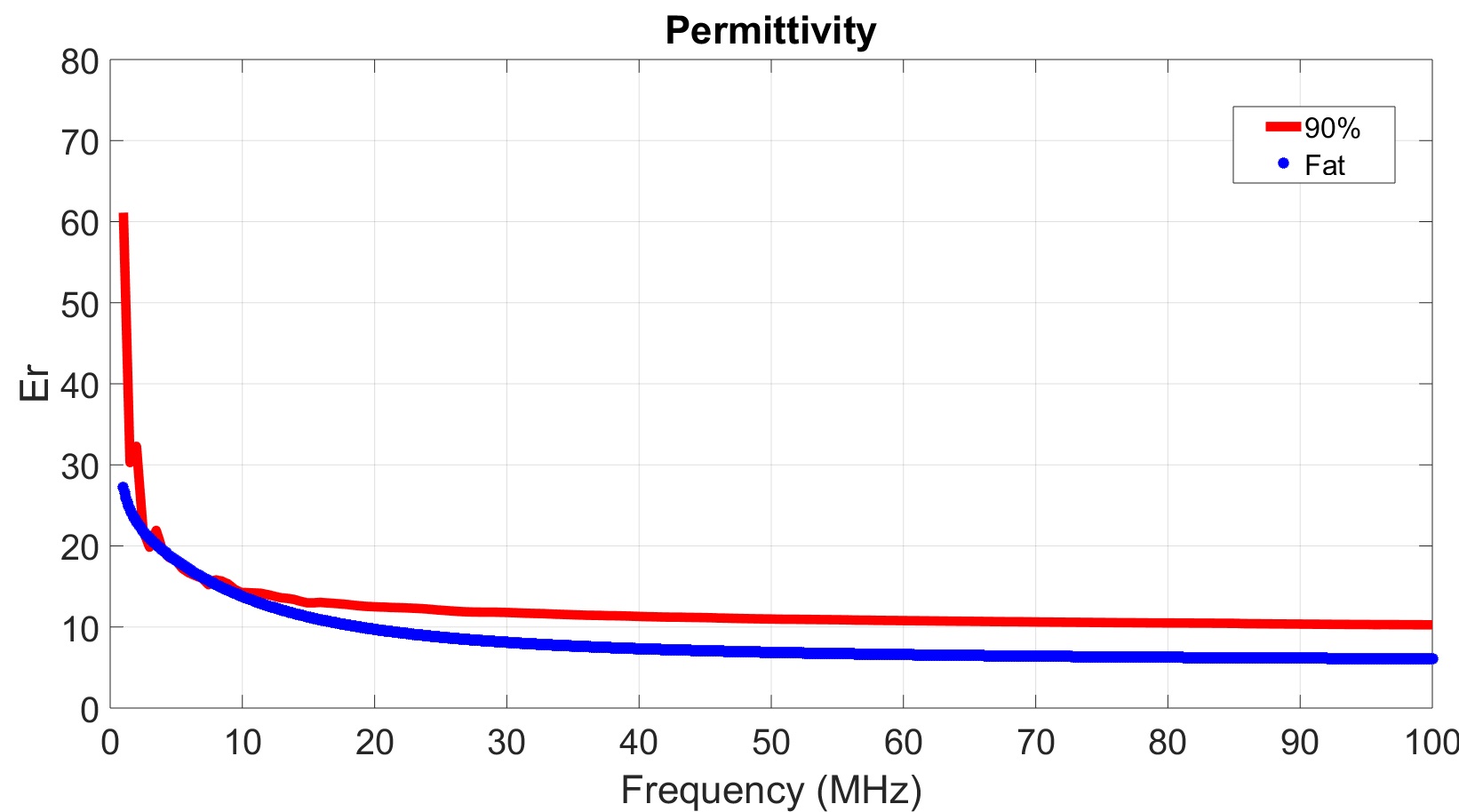} }}%
   
      \caption{Matching tissues with different samples that show good accuracy withing certain frequency ranges; (a) 80\% oil-kerosene with the cortical bone, from the permittivity point of view, and specially for frequencies greater than 30 MHz, (b) 80\% oil-only solution with the bone marrow, from the conductivity point of view, within the  frequency range 12 MHz till 100 MHz, (c) (b) 80\% oil-only solution for the Fat tissue,from the permittivity point of view, specially for low frequencies.}%
      \label{fig6_eighty_percent}%
  \end{figure}
  %\begin{figure}[]
 %      \centering
 %\includegraphics[height=2.5in,width=3.28in,scale=1]{Figure-6_Khorshid.jpg}
 %\caption{An example for matching the right sample with the exact oil solution concentration with the tissue of concern, where the 80\% oil-kerosene solution showed an accurate match with the cortical bone, from the permittivity point of view, and specially for frequencies greater than 30 MHz.}
 %\label{fig6_eighty_percent}
  %  \end{figure}
  \begin{figure}[H]
       \centering
 \includegraphics[height=2.1in,width=1.9in,scale=1]{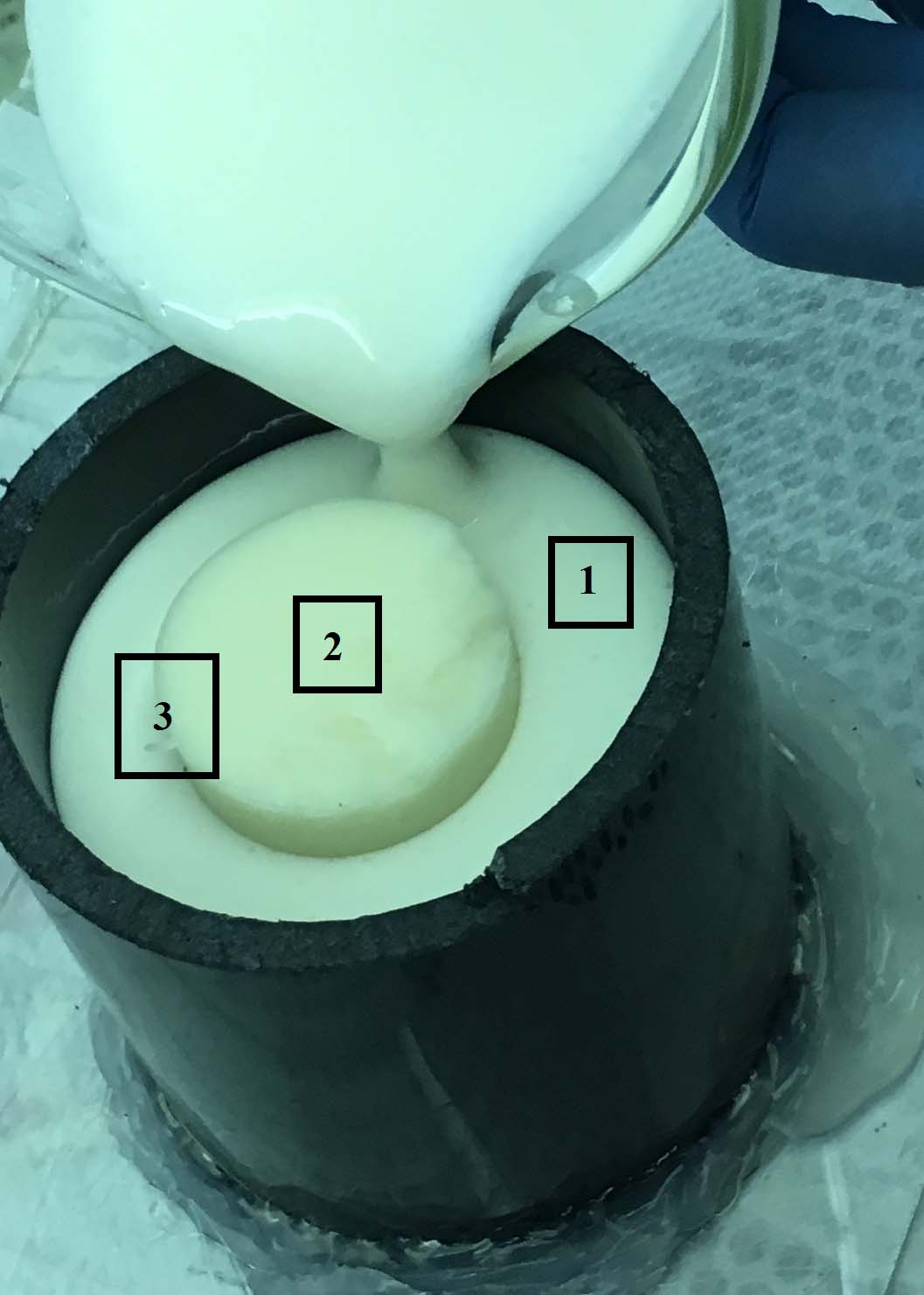}
 \caption{The oil-kerosene composite sample being prepared, where the 60\% concentration sample is being poured into a mold containing an already semi-solid 20\% concentration sample. Inscribe are the positions where samples were taken for experimental testing; 1-from the 60\% sample,2- from the 20\% sample and finally 3 is take from the interface at both samples.}
 \label{fig7_composite}
    \end{figure}
\section{Composite Sample}
Building a composite sample that spans multiple layers is essential to observe the behavior of the electric signal and how it diffuses from the skin into other layers while propagating  
from one node to the other. Towards that goal, a composite oil-kerosene sample was prepared. First, an oil-kerosene sample with 20\% oil solution concentration was prepared, and then left it in the mold for two days, to allow the sample to solidify and the chemical reactions to conclude. The sample was then placed in a bigger mold, where a 60\% oil-kerosene sample was prepared and poured in the same mold, forming a composite sample in the form of two concentric cylinders, with the 20\% sample enclosed by the 60\% one, as shown in Figure \ref{fig7_composite}. The composite sample was then kept at room temperature for another two days, before experimental measurements were executed. Samples were then taken from three different positions to measure the electrical properties, as shown in Figure \ref{fig7_composite}; samples were first taken from position 1 which is deeply in the 60\% concentration sample section, others taken from position 2 which is in the middle of the 20\% concentration sample, and finally the last samples were taken from position 3 which is at the interface between both samples. Results were averaged over the samples taken at each position and plotted in Figure \ref{fig8_dielectric_properties_composite}. It is clear that the permittivity for both concentrations remained almost the same; however the 60\% sample was a more affected at the interface than the 20\% one. Similarly, the conductivities for the same samples are plotted in Figure \ref{fig8_dielectric_properties_composite} (b). The conductivity values at the interface were more affected by depositing two different concentrations side by side compared to the permittivity ones. However, both curves, when compared with the results in Figure \ref{fig4_dielectric_properties_oil_only} and \ref{fig5_dielectric_properties_oil_kerosene}, show that the electrical properties at the core of each sample is almost unchanged, which proves the capability of these materials in the preparation of a composite multi-layered phantom models, for an accurate mimicking of the human body, from the electrical properties point of view.

\begin{table*}[t]
\centering
\caption{Matching Tissues With Samples of Best Accuracy (less than 10\% matching error) for Different Frequency Ranges  }
\label{table3}
%\begin{minipage}{9cm}
\begin{tabular}{l|l|l|l|l|l|l|}
\cline{2-7}
                                                     & \multicolumn{6}{c|}{\textbf{Electrical Property}}                                                                                           \\ \cline{2-7} 
                                                     & \multicolumn{3}{l|}{\textbf{Conductivity}}                          & \multicolumn{3}{l|}{\textbf{Permittivity}}                          \\ \hline
\multicolumn{1}{|l|}{\textbf{Tissue}}                & \textbf{Sample Concentration} & \textbf{Fmin (MHz)} & \textbf{Fmax(MHz)} & \textbf{Sample Concentration} & \textbf{Fmin} & \textbf{Fmax} \\ \hline
\multicolumn{1}{|l|}{\multirow{3}{*}{Cortical Bone}} & 70\% (Oil-Kerosene)           & 4.2              & 11               & 60\% (Oil-Kerosene)           & 1.8              & 7                \\ \cline{2-7} 
\multicolumn{1}{|l|}{}                               & 60\% (Oil Only)               & 5.9              & 100              & 70\% (Oil-Kerosene)           & 11.8             & 20               \\ \cline{2-7} 
\multicolumn{1}{|l|}{}                               & -                             & -                & -                & 80\% (Oil-Kerosene)           & 30               & 100              \\ \hline
\multicolumn{1}{|l|}{Bone Marrow}                    & 80\% (Oil Only)               & 12.8             & 100              & 90\% (Oil Only)               & 1.8              & 25               \\ \hline
\multicolumn{1}{|l|}{\multirow{4}{*}{Dry Skin}}      & 30\% (Oil Only)               & 7                & 9                & 10\% (Oil Only)               & 25               & 33.7             \\ \cline{2-7} 
\multicolumn{1}{|l|}{}                               & 20\% (Oil Only)               & 10               & 14.5             & 10\% (Oil Only)               & 42               & 58               \\ \cline{2-7} 
\multicolumn{1}{|l|}{}                               & -                             & -                & -                & 20\% (Oil Only)               & 58.4             & 90               \\ \cline{2-7} 
\multicolumn{1}{|l|}{}                               & -                             & -                & -                & 30\% (Oil Only)               & 93               & 100              \\ \hline
\multicolumn{1}{|l|}{\multirow{4}{*}{Wet Skin}}      & -                             & -                & -                & 10\% (Oil Only)               & 11.9             & 16.8             \\ \cline{2-7} 
\multicolumn{1}{|l|}{}                               & -                             & -                & -                & 20\% (Oil-Kerosene)           & 24               & 38               \\ \cline{2-7} 
\multicolumn{1}{|l|}{}                               & -                             & -                & -                & 30\% (Oil Only)               & 38               & 71               \\ \cline{2-7} 
\multicolumn{1}{|l|}{}                               & -                             & -                & -                & 40\% (Oil Only)               & 73               & 100              \\ \hline
\multicolumn{1}{|l|}{Fat}                            & 80\% (Oil-Kerosene)           & 11               & 100              & 90\% (Oil Only)               & 2.3              & 11.5             \\ \hline
\multicolumn{1}{|l|}{\multirow{2}{*}{Muscle}}        & -                             & -                & -                & 30\% (Oil Only)               & 24               & 54               \\ \cline{2-7} 
\multicolumn{1}{|l|}{}                               & -                             & -                & -                & 30\% (Oil-Kerosene)           & 39               & 100              \\ \hline

\end{tabular}
%\end{minipage}
\end{table*}
\section{Conclusion}
In this paper different materials were investigated to propose a reproducible, accurate and reliable tissue mimicking materials, to be used in constructing a multilayer phantom that 
can serve as a test platform for intra-body communication applications. Prior trials in the IBC field were shown, where phantoms were simply considered as a one homogeneous layer, and are mostly liquid phantoms, which eliminate the possibility of using any of them in constructing a multilayer phantom. Two methods were considered for preparing tissue mimicking samples, one utilizing oil together with other components, while the other method mixes oil with kerosene, before adding other ingredients. Different samples with different oil solution percentages were prepared and tested, to measure their electrical properties and find the most adequate for mimicking different tissues, which would depend on the application in hand; whether conductivity or permittivity is of more concern, as well as the range of frequencies of interest. A composite sample was then prepared, where a sample of 60\% oil solution concentration was deposited around another of 20\% concentration, forming a multilayer model in the form of concentric cylinders. Dielectric properties of the resulting sample were then tested and compared with various regions across the sample to study the affect of layers deposition, showing how the interface between both layers causes slight changes in the samples' properties, yet maintaining the properties within each layer separately. The results show the efficacy of using such materials in constructing a multilayer phantom mimicking the dielectric properties for the layers of interest over the range of frequencies of concern for IBC application.

  \begin{figure}[]%
      \centering
      \subfloat[]{{\includegraphics[height=2.3in,width=3.4in]{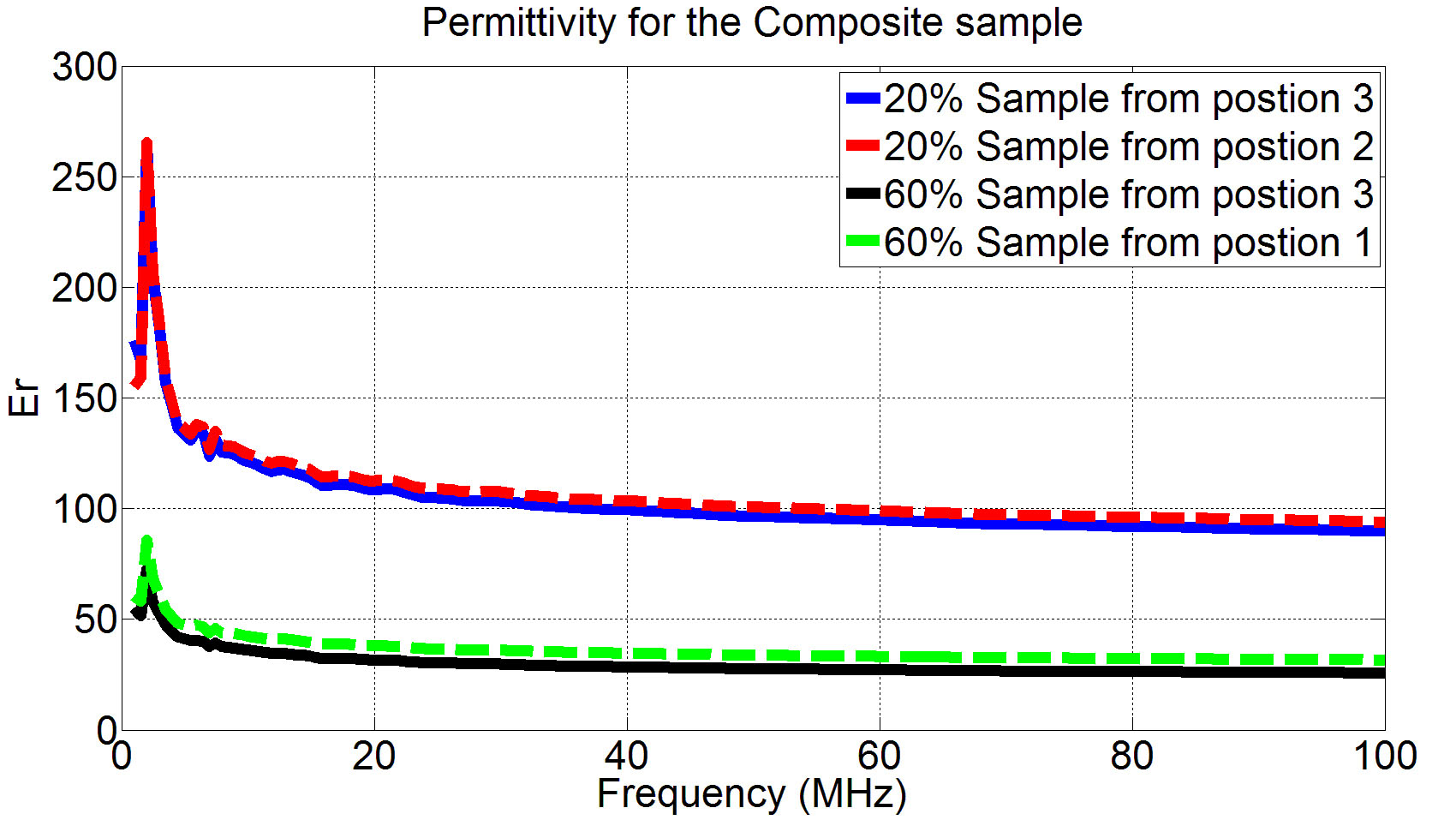} }}%
    \qquad  
      \subfloat[]{{\includegraphics[height=2.3in,width=3.4in]{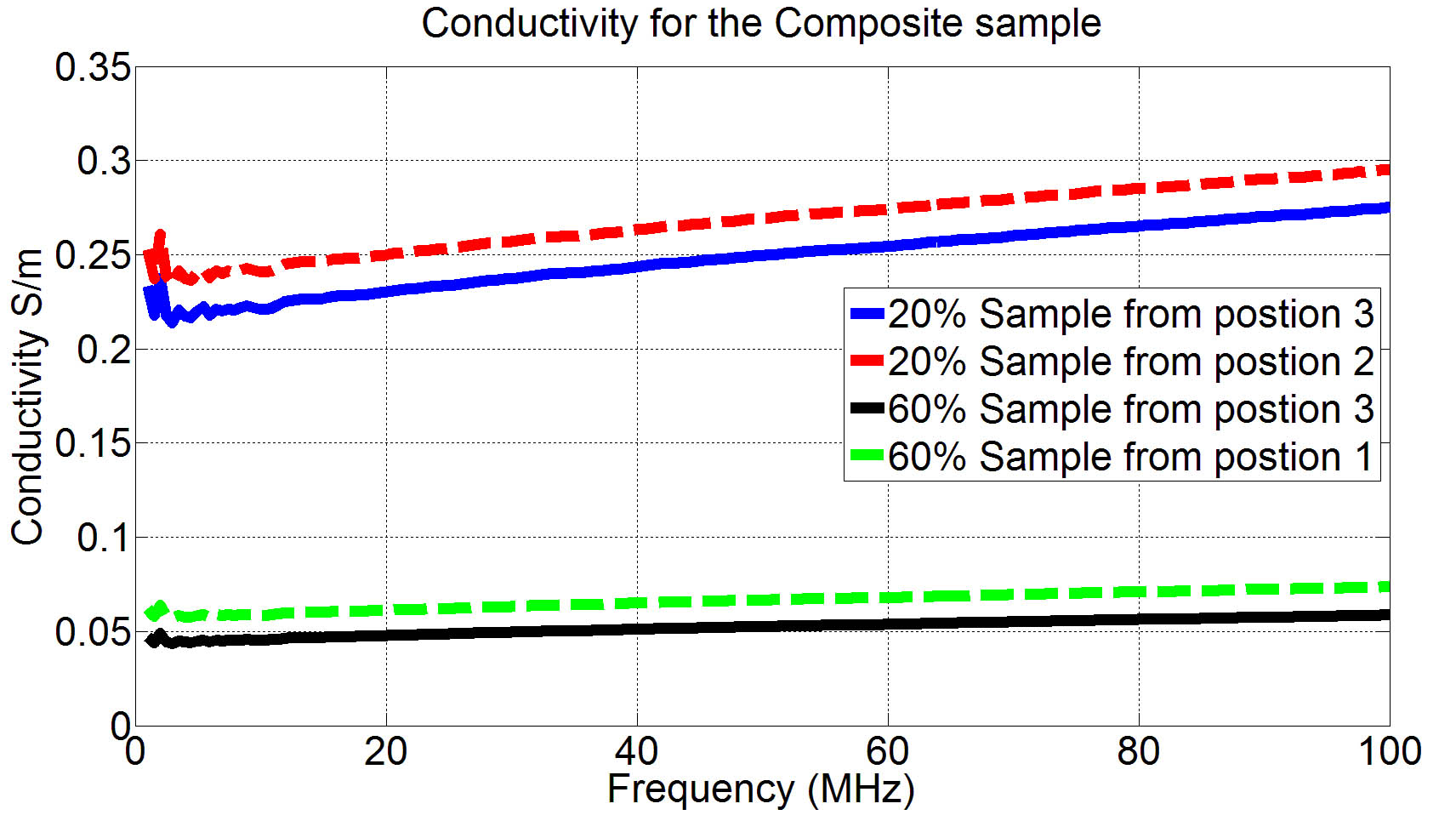} }}%
      \caption{(a) Measured permittivity for samples taken from positions 1, 2 and 3 as indicated in Fig.6, (b) Measured conductivity for the same samples.}%
      \label{fig8_dielectric_properties_composite}%
  \end{figure}
\section*{Acknowledgements}
\par
The authors thank Professor G. P. Li, Professor Mark Bachman and Sarkis Babikian at the Micro Integrated Devices and Systems Laboratory, University of California, Irvine for their help and guidance concerning preparing the samples and the measurement setup.

\begin{IEEEbiography}[{\includegraphics[width=1.1in,height=1.22in]{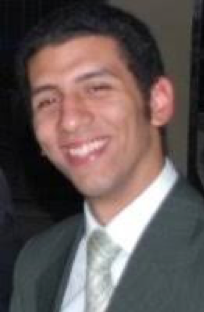}}]{Ahmed E. Khorshid}  received the B.Eng. degree (Hons.) in Electronics and Communications Engineering
and the M.Sc. degree in Electronics Engineering from the Faculty of Engineering, Cairo University, Cairo,
Egypt, in 2010 and 2013, respectively. He is currently pursuing the Ph.D. degree with the University of
California, Irvine, USA. His research interests include body area networks, IoT applications, Wearbles,
healthcare systems, machine learning, and analog mixed circuits. He received several awards and
fellowships including the Broadcom Foundation Fellowship and the National Institute of Justice
Graduate Research Fellowship in Science, Technology, Engineering, and Mathematics.
\end{IEEEbiography}
\begin{IEEEbiography}[{\includegraphics[width=1.1in,height=1.22in]{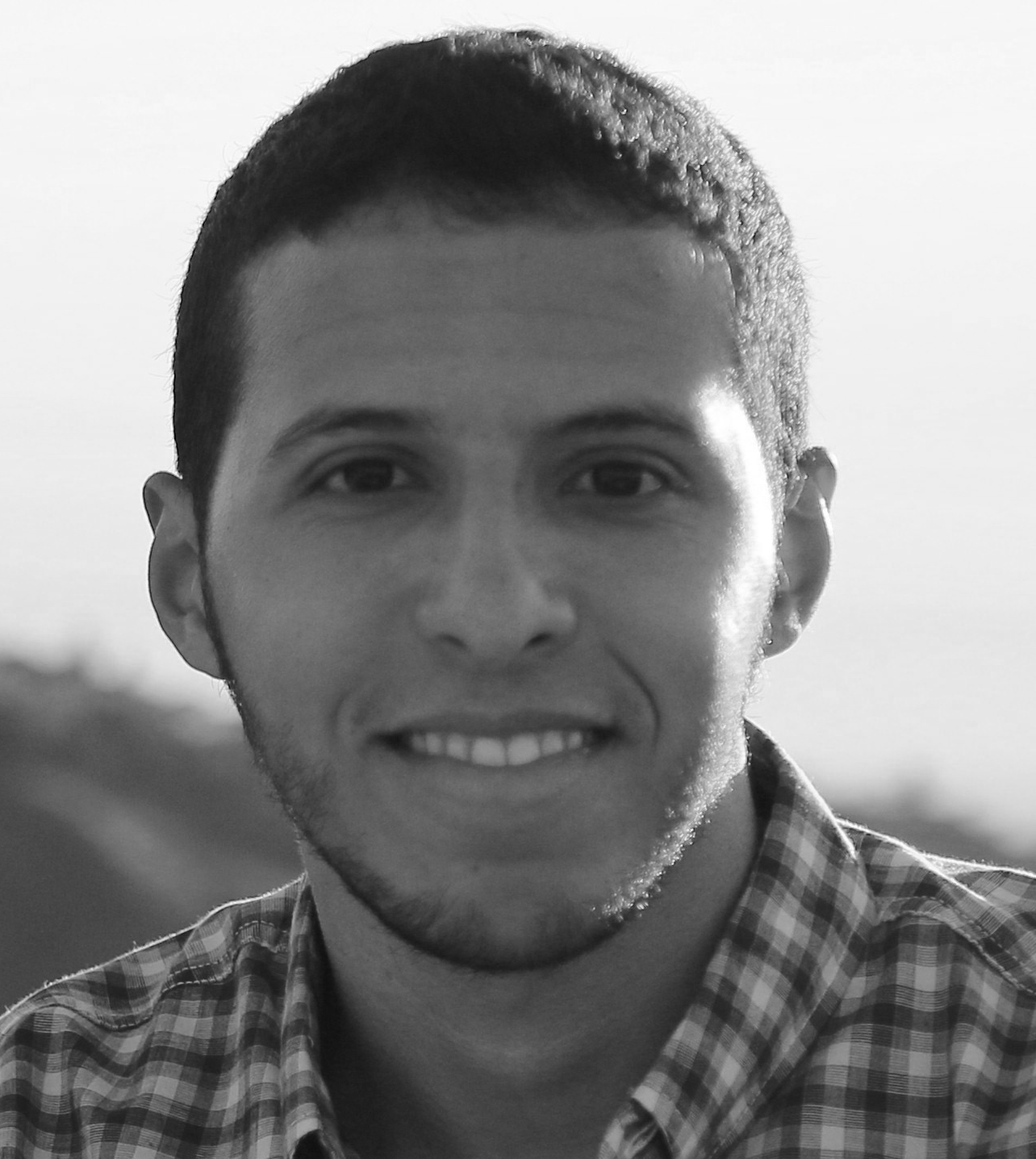}}]{Ibrahim N. Alquaydheb}  received the B.S. degree in electrical engineering (with honors) from Prince Sattam bin Abdulaziz University in 2013. He is currently working toward his M.S. degree at University of California, Irvine. His research is focused on body area networks, biometric authentication systems, wearables and ultra-low power wireless systems. He worked as an electrical engineer at the Arabian BEMCO Contracting Company. In 2014, he was a teaching assistant at Prince Sattam bin Abdulaziz University.     
\end{IEEEbiography}
\begin{IEEEbiography}[{\includegraphics[width=1.05in,height=1.22in]{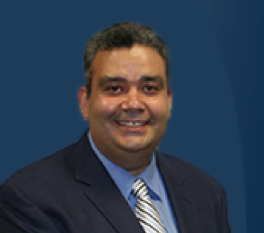}}]{Ahmed M. Eltawil}  (S'97--M'03--SM'14) is a Professor at the University of California, Irvine. He has been with the Department of Electrical Engineering and Computer Science since 2005 where he is the founder and director of the
Wireless Systems and Circuits Laboratory. His current research interests are in the general area of low power digital
circuit and signal processing architectures with an emphasis on mobile computing and communication systems. In
addition to his department affiliation, he is also affiliated to a number of research centers across the University of
California, Irvine. He received the Doctorate degree from the University of California, Los Angeles, in 2003 and the
M.Sc. and B.Sc. degrees (with honors) from Cairo University, Giza, Egypt, in 1999 and 1997, respectively. Dr.
Eltawil has been on the technical program committees and steering committees for numerous workshops, symposia,
and conferences in the areas of low power computing and wireless communication system design. He received
several awards, as well as distinguished grants, including the NSF CAREER grant in 2010 supporting his research in
low power systems. In 2015, Dr Eltawil founded Lextrum Inc., to develop and commercialize full duplex solutions
for 5G communications systems.
\end{IEEEbiography}
\begin{IEEEbiography}[{\includegraphics[width=1.1in,height=1.22in]{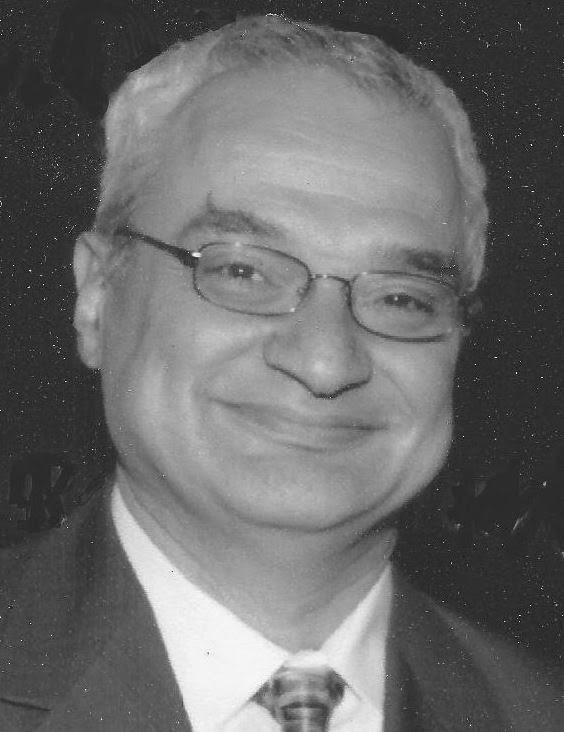}}]{Fadi J. Kurdahi}  (S'85-M'87-SM'03-F'05)  received the B.E. degree in electrical engineering from the American University of Beirut in 1981 and the Ph.D. degree from the University of Southern California in 1987. Since then, he has been a Faculty with the Department of Electrical Engineering and Computer Science, University of California at Irvine, where he conducts research in the areas of computer aided design, high level synthesis, and design methodology of large scale systems, and serves as the Director of the Center for Embedded \& Cyber-physical Systems, comprised of world-class researchers in the general area of Embedded and Cyber-Physical Systems. He is a fellow of the AAAS. He was the Program Chair or the General Chair on program committees of several workshops, symposia, and conferences in the area of CAD, VLSI, and system design, and served  on editorial boards of several journals. He received the best paper award of the IEEE TRANSACTIONS ON VLSI in 2002, the best paper award in 2006 at ISQED, and four other distinguished paper awards at DAC, EuroDAC, ASP-DAC, and ISQED. He also received the Distinguished Alumnus Award from his Alma Mater, the American University of Beirut, in 2008.
\end{IEEEbiography}

\EOD

\end{document}